\documentclass[%
 reprint,
nofootinbib,
 amsmath,amssymb,
 aps,
]{revtex4-2}

\usepackage{graphicx}% Include figure files
\usepackage{dcolumn}% Align table columns on decimal point
\usepackage{bm}% bold math
\usepackage{aas_macros}
\usepackage{color}

\newcommand{\mrm}[1]{_{\rm #1}}

\begin{document}

\begin{flushright}
CERN-TH-2020-067
\end{flushright}

\preprint{APS/123-QED}

\title{Stellar Signatures of Inhomogeneous Big Bang Nucleosynthesis}

\author{Alexandre Arbey}
\email{alexandre.arbey@ens-lyon.fr}
\affiliation{Univ Lyon, Univ Claude Bernard Lyon 1, CNRS/IN2P3, IP2I Lyon, UMR 5822, F-69622, Villeurbanne, France}%
\affiliation{Theoretical Physics Department, CERN, CH-1211 Geneva 23, Switzerland}%
\affiliation{Institut Universitaire de France (IUF), 103 boulevard Saint-Michel, 75005 Paris, France}

\author{J\'er\'emy Auffinger}
    \email{j.auffinger@ipnl.in2p3.fr}
    \affiliation{Univ Lyon, Univ Claude Bernard Lyon 1, CNRS/IN2P3, IP2I Lyon, UMR 5822, F-69622, Villeurbanne, France}%

\author{Joseph Silk}
\email{joseph.silk@physics.ox.ac.uk}
\affiliation{Sorbonne Universit\'e, CNRS, UMR 7095, Institut d’Astrophysique de Paris, 98 bis bd Arago, 75014 Paris, France}%
\affiliation{Department of Physics and Astronomy, Johns Hopkins University, Baltimore MD 2218, USA}
\affiliation{Beecroft Institute of Particle Astrophysics and Cosmology, University of Oxford, Oxford OX14BN, UK}

\date{\today}% It is always \today, today,

\begin{abstract}
    We evaluate abundance anomalies generated in patches of the universe where the baryon-to-photon ratio was locally enhanced by possibly many orders of magnitude in the range $\eta = 10^{-10} - 10^{-1}$. Our study is motivated by the possible survival of rare dense regions in the early universe, the most extreme of which, above a critical threshold, collapsed to form primordial black holes. If this occurred, one may expect there to also be a significant population of early-forming stars that formed in similar but subthreshold patches. We derive a range of element abundance signatures by performing BBN simulations at high values of the baryon-to-photon ratio that may be detectable in any surviving first generation stars of around a solar mass. Our predictions apply to metal-poor  galactic halo stars, to old globular star clusters and to dwarf galaxies, and we compare with observations in each of these cases.
\end{abstract}

\maketitle

%\tableofcontents

\section{Introduction}
\label{sec:introduction}

Big bang nucleosynthesis (BBN) is a fundamental probe of the first few minutes of the universe. Strong constraints are set on the physics of the early universe \cite{2018PhRvD..98c0001T}. Both the baryon density  and possible deviations in the number of relativistic species from the standard value  are severely constrained, especially in the light of the Planck data on Cosmic Microwave Background (CMB) fluctuations \cite{fields}. There are however persistent anomalies, including the primordial lithium abundance \cite{iocco} and possible indications of helium and other light element anomalies that are presumably due   to as yet unresolved issues in stellar evolution modelling \cite{2018Ap.....61..262L}.

It behoves us to carefully test the homogeneity of BBN, which is one of the fundamental assumptions of Standard BBN (SBBN). Here we evaluate abundance anomalies generated in patches of the universe where the baryon-to-photon ratio was locally enhanced by possibly many orders of magnitude. Our study is motivated by the possible survival of rare dense regions in the early universe that collapsed to form primordial black holes (PBH), a subject of intense current scrutiny for its role in contributing to the dark matter abundance and/or seeding early supermassive black hole formation. Such regions are commonly thought to derive from primordial isocurvature perturbations, where the initial baryon-to-photon ratio is considerably enhanced relative to the standard model. These overdense regions might have later formed stars that retained a memory of inhomogeneous BBN. 

Here we explore further the inhomogenous BBN signatures for extreme values of $\eta = 10^{-10} - 10^{-1}$ motivated by primordial black hole formation, on the grounds that fluctuations below the density threshold on the horizon scale  to form a PBH \cite{musco} could still form rare stars, or even antistars, at very early epochs. We limit ourselves to baryon density fluctuations with positive $\eta$ over a very broad range. Previous studies, following the pioneering work by Wagoner \textit{et al.} \cite{Wagoner1967}, were limited to probing only a restricted range in $\eta \lesssim 10^{-3}$ \cite{2004PThPh.112..971M,Nakamura2010,Nakamura2013}. We  will demonstrate that exotic abundance signatures, most notably produced by rapid neutron capture in rare core-collapse supernovae (SNe) or binary neutron star mergers \cite{Komiya2016,Duggan2018,2019arXiv190101410C,Cote2019}, may also contain a possible tracer of inhomogeneous BBN as predicted in \cite{Applegate1988}. These $r$-process abundances are observed to be enhanced in metal-poor halo stellar populations \cite{2018ApJ...858...92H} and in metal-poor dwarf galaxies \cite{2019ARA&A..57..375S}, where we argue that exotic BBN signatures may also be hidden. Our paper is organized as follows: Section \ref{sec:context} presents the context of the study, Section \ref{sec:numerics} describes the nuclear reaction chains and the numerical methods used, Section \ref{sec:comparisons} presents the results of the computations and the comparison with data, and we conclude in Section \ref{sec:conclusion}.

\section{Context}
\label{sec:context}

Rare patches of the universe may have initially had extremely high baryon-to-photon ratios. This is motivated by the possibility that rare large amplitude primordial isocurvature perturbations with an extremely blue spectrum and with residual power on stellar mass-scales were present in the very early universe. These fluctuations may have been responsible for PBH formation over a variety of horizon masses, the natural scale for PBH formation. Scales of interest range  from supermassive black holes (SMBH), intermediate mass black holes (IMBH), stellar mass black holes and substellar to sublunar mass black holes. 

These ideas have attracted recent attention as a potential source of primordial black holes for the following reasons. Such arguments are motivated by LIGO/VIRGO observations, which admittedly may have a more conventional astrophysical black hole origin, but are testable. These tests consist of searching for solar mass black hole binary mergers, black holes mergers in the pair instability gap above $\sim 60\, \rm M_\odot,$ and black hole mergers at high redshift where PBHs should be more abundant than astrophysical BHs. 
However increasingly stringent limits, mostly from microlensing experiments, restrict the mass fraction of PBHs in the stellar mass range to be at most $\sim 1\%$ of the dark matter. 

Attention of theorists consequently shifted  towards two complementary directions. 
Firstly, the subdominant nature of massive PBHs allows the possibilty that they still could be present to solve other urgent astrophysical problems.
Extreme isocurvature fluctuations, admittedly on somewhat smaller scales, provide, for example, the primordial IMBHs needed to seed the observed SMBHs
\cite{inayoshi}, 
Indeed some (or even all) SMBHs themselves could plausibly be primordial and would have naturally formed at the BBN epoch, when the horizon enclosed $\sim 10^9\,\rm M_\odot.$ 

Secondly, the dark matter window for PBHs remains open at much lower PBH masses. The inhomogeneous regions that we invoke here could equally be the tail of the  mass distribution that generated  much smaller PBHs (of sublunar mass) at a correspondingly earlier epoch. These are potential, and indeed the only viable, PBH dark matter candidates 
(see \cite{carr19} and references therein for a recent review on PBH formation mechanisms and constraints).
 
The PBH  hypothesis has an important consequence. For any plausible initial conditions, one is likely to have many uncollapsed isocurvature fluctuations (``failed'' PBHs). These will have late epoch signatures. One is gravitational wave production, possibly observable as a stochastic background 
\cite{nakama}. 
A second consequence is that there are implications for
the epoch of BBN, when the universe may be inhomogeneous  on  scales comparable to those of the hypothesized PBH masses, albeit admittedly in rare patches.  

However such patches are rare for two reasons: firstly standard BBN is a great  success and limits on inhomogeneity on horizon patches are $\lesssim 17\% $, expressed in terms of  $\Delta\eta$  
%\Delta Y_p\sim 0.01.$
\cite{barrow}.
Secondly, the universe at BBN epoch is highly radiation-dominated, and the associated PBHs must be subdominant by a factor of at least $(1+z_{\rm BBN})/(1+z_{\rm eq})\sim 10^4$
\cite{carr18}.

Now BBN of neutron-capture elements in high $\eta$ patches occurred at an epoch $t\sim 10^3$\,s  when the baryon content of the horizon is around $10^4\,\rm M_\odot.$ This gives the natural (minimum) scale of the patches of high $\eta$ that we  hypothesise as inhomogeneous sites of BBN. It is not a large step to imagine that such inhomogeneities may occur on the scales of the smallest dwarf galaxies or star clusters at the epoch of first star formation,  once H$_2$ cooling develops at $z\sim 10-20.$ These are the scales on which we may search for signatures of inhomogeneous BBN.

Let us  define the degree of inhomogeneity that we will consider. At BBN, the required baryon overdensity of an isocurvature fluctuation is at least $ (1+z_{\rm BBN})/(1+z_{\rm eq})\sim 10^5\,\eta_0$ where the standard (Planck) value of the baryon-to-photon ratio is $\eta_0 = (6.104\pm 0.058)\times 10^{-10}.$ In this paper, we have explored the range  $\eta = 1-10^9\,\eta_0$. We will show  that the most interesting region for possible abundance signatures is at relatively  high $\eta$, a range that  has not previously been explored in any detail.

Should evidence  for stellar mass PBHs be confirmed, we will argue that there is a strong case for searching for stellar relics that are in effect failed PBHs. These can be distinguished from Population III stellar survivors by the unique abundance signatures that we find below. 
We note that recent simulations of the mass function of Population III, long considered to be massive stars of $\sim 100-1000\rm\, M_\odot$, demonstrate that ongoing fragmentation to below a solar mass can indeed occur \cite{stacy}.

Any such imprints will also result in inhomogeneous BBN, surviving on similar mass scales that have not collapsed -- being subthreshold -- but would be greatly diluted by the present epoch because of mass mixing on galactic scales, stellar evolution, stellar mass loss and baryonic circulation in the interstellar medium. However initial signals could still survive as anomalous primordial abundances and be visible in a small fraction of the oldest stars, in the galactic halo and in the oldest, most metal-poor dwarf galaxies.

The baryon-to-photon ratio $\eta\equiv n\mrm{b}/n_\gamma$, while tightly constrained by the CMB data in homogeneous BBN to be $\eta=(6.104\pm 0.058)\times 10^{-10}$ \cite{2020JCAP...03..010F}, can take on a wide range of values. Signatures of a high-$\eta$ value include elevated helium abundances and trace amounts of exotic elements normally produced in stars but possibly overproduced in nonstandard BBN. Rare patches of inhomogeneity in $\eta$ at the epoch of BBN can have late-time stellar signatures once mixing occurs on galactic scales. Recall that the baryon content within  the particle horizon  at the onset of BBN is only $\sim 100\, \rm M_\odot$. If sufficiently anomalous in terms of rare elements produced by extremely high $\eta$ BBN is localized to rare patches, late formation of stars may reflect contamination by unusual abundance patterns.

A prime motivation for  such extremely inhomogeneous BBN comes from the fact that the existence of PBHs is most plausibly explained by extreme but rare  isocurvature fluctuations  generated  in an Affleck-Dine-like early phase transition  associated with baryogenesis \cite{affleck}. Such an early phase transition  even allows the sign of the baryon-to-photon ratio to be inverted. Indeed PBHs have  formed in patches with positive or negative baryon number. An extreme signature of such an effect would be provided by inhomogeneous BBN in patches where $\eta$ changes sign. If they survived to the present epoch, such patches of antimatter could later form antistars, distinguishable by their abundance patterns as contaminated by the unusual local BBN history. Our BBN predictions are independent of the sign of $\eta$, and apply equally to any relic antistars.

Antistars have long been advocated by Dolgov and collaborators \cite{1993PhRvD..47.4244D,2009NuPhB.807..229D,2015PhRvD..92b3516B} in the context of baryon number fluctuations generated by an early universe first order phase transition. Arguments based on measurement of cosmic ray antihelium nuclei by AMS-02 have recently been put forward \cite{2019PhRvD..99b3016P} to motivate such a scenario.

\section{Numerical methods}
\label{sec:numerics}

When the baryon-to-photon ratio reaches high values, we expect heavy elements to form in appreciable quantities compared to SBBN where only the light elements have a detectable final abundance. As we do not know a priori at which atomic number $A$ we should stop the nuclear grid, we have decided to consider the full REACLIB grid \cite{REACLIB} for the nuclear chains. We use the Big Bang Nucleosynthesis public code \texttt{AlterBBN} \cite{AlterBBN1,AlterBBN2} which allows the users to compute the abundances of the elements in standard and alternative cosmological scenarios and which has been recently updated -- e.g. to include the REACLIB database for nuclear reactions and to reach $\eta \sim \mathcal{O}(1)$ with an acceptable precision.

We have performed a first \texttt{AlterBBN} computation for $\eta = 10^{-15} - 10^{1}$ using the full nuclear grid. In this way we have identified the nuclei that do not play a significant role in the BBN evolution at large $\eta$, and we hence removed from the grid all nuclei whose abundance relative to hydrogen did not exceed a lower cut-off of $10^{-12}$ at any time during BBN computation and for all $\eta$ values in the considered range. In this way we reduced the number of nuclei from several thousands to $\sim 800$. We also checked that the abundances of these $\sim 800$ nuclei where unchanged when performing a BBN computation with this subset of nuclei. Then we identified in the remaining nuclei those for which the abundance is at least $[X]/[H] \gtrsim 10^{-10}$ for some value of $\eta$. This led us to a set of $132$ nuclei (corresponding to $54$ chemical elements) for which we present the abundances here\footnote{The full sets of data -- both reduced and not reduced -- in the form of numerical tables and a pdf file containing plots of the reduced data are provided as supplemental material to this article.}. The numerical methods needed to compute the abundances of such a great number of nuclei -- from hydrogen to $^{250}$Cf -- converge slowly (see \cite{AlterBBN2} for details about these methods). In addition the results for $\eta \gtrsim 0.1$ are difficult to interpret because of the unclear numerical errors, and in the following we therefore limit ourselves to $\eta < 0.1$.

\section{Results and comparisons with observations}
\label{sec:comparisons}

\subsection{Comparison with solar abundances}

\begin{figure*}[t!]
    \centering
    \includegraphics[scale = 1]{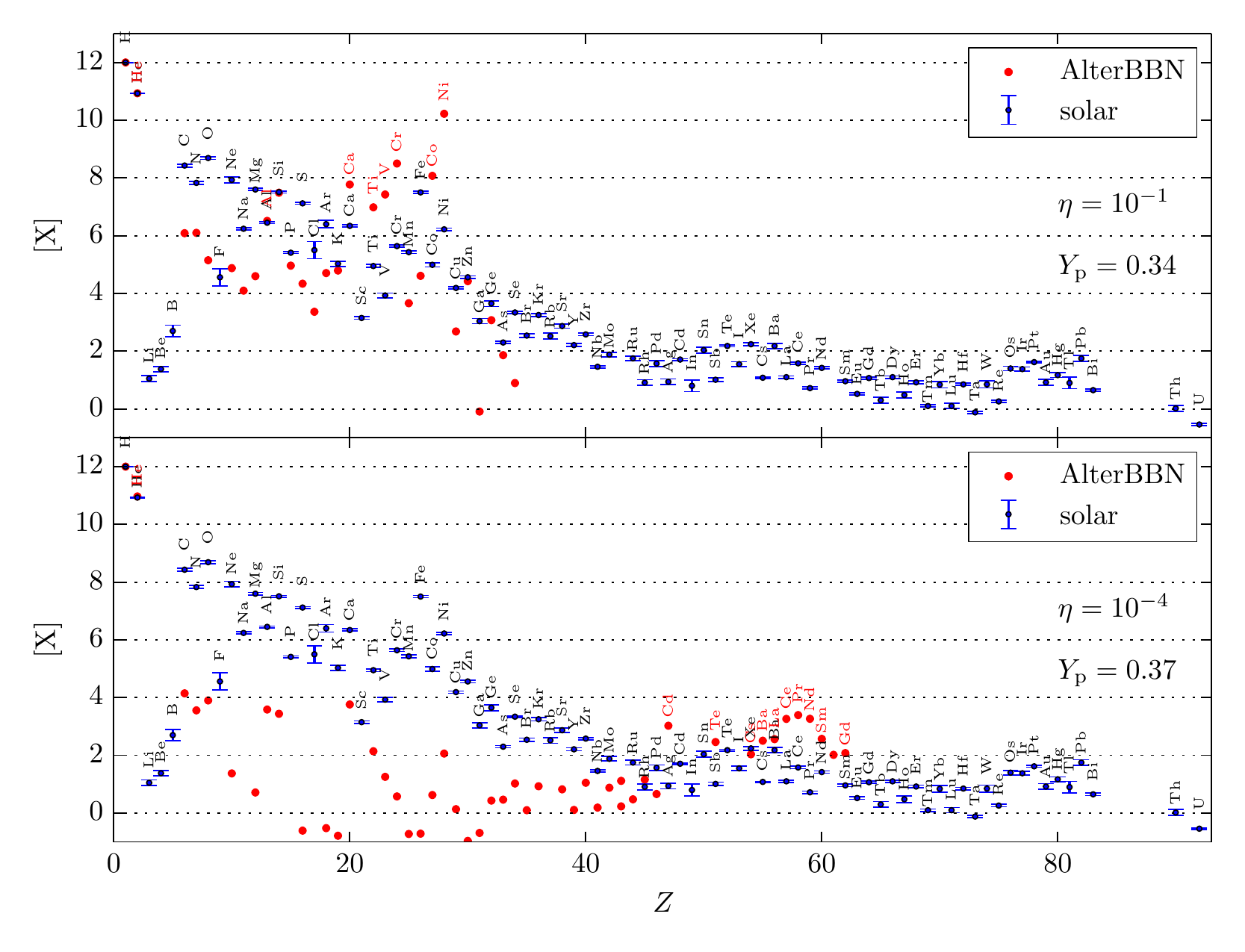}
    \caption{Abundances of the elements for fixed values of $\eta = \{10^{-1},10^{-4}\}$ corresponding to $Y\mrm{p} = \{0.34,0.37\}$ computed by \texttt{AlterBBN} (red dots) and compared to solar abundances of \cite{Asplund2009} (blue dots, when available). The names of overproduced elements compared to solar abundances are written in red. See text for the definition of the [X] notation.}
    \label{fig:fixed_eta}
\end{figure*}

First we compare the data obtained with \texttt{AlterBBN} at high-$\eta$ with the solar abundances of Asplund \textit{et al.} \cite{Asplund2009}. When $\eta$ is larger than the SBBN value $\eta_0 \sim 6\times10^{-10}$, we find that groups of elements in addition to helium are overproduced compared to the solar abundances. Throughout this paper, abundances are given in the standard stellar physics form, if we consider element X then
\begin{equation}
    {\rm [X]} \equiv \log_{10}\left(\dfrac{n({\rm X})}{n({\rm H})}\right) + 12\,,
\end{equation}
where $n({\rm X})$ (respectively $n({\rm H})$) is the number density of X nuclei (respectively hydrogen $^1$H). Fig.~\ref{fig:fixed_eta} shows that when $\eta = 10^{-4}$, the overproduced elements have $Z \sim 60$, while when $\eta = 10^{-1}$ the overproduction concerns $Z \sim 25$, with a continuous evolution between these extreme values. When $\eta$ is smaller than $10^{-5}$ no heavy elements ($Z \gtrsim 10$) are produced. These results are comparable to those of Refs.~\cite{2004PThPh.112..971M,Nakamura2010,Nakamura2013}, even if in the latter references the computations were limited to $\eta \lesssim 10^{-3}$ and used a less complete set of nuclei. The shift between high-$Z$ $r$-process elements for $\eta \lesssim 10^{-3}$ and lower $Z$ $p$-process elements for $\eta \gtrsim 10^{-3}$ is described in \cite{2004PThPh.112..971M} and results from an efficient active proton capture that prevents synthesis of heavier elements due to Coulomb barrier effects. The competition between the expansion rate that is modified in high-$\eta$ regions due to local matter domination and the nuclear rate can also contribute to this shift.

Our purpose is now to compare our predicted high $\eta$ signatures of inhomogeneous BBN with the abundance signatures in the oldest stellar systems, where the first generations of stars may have retained anomalies even if greatly diluted by mixing at dwarf galaxy scales. As we discuss below, mixing was incomplete in $r$-process enhanced ultra metal-poor dwarf spheroidals such as Reticulum II and Tucana III, and signatures of inhomogeneous BBN might conceivably have survived. 

\subsection{High Helium abundance and globular clusters}
\label{sec:helium}

\begin{figure*}[t!]
	\centering{
		\includegraphics[scale = 1]{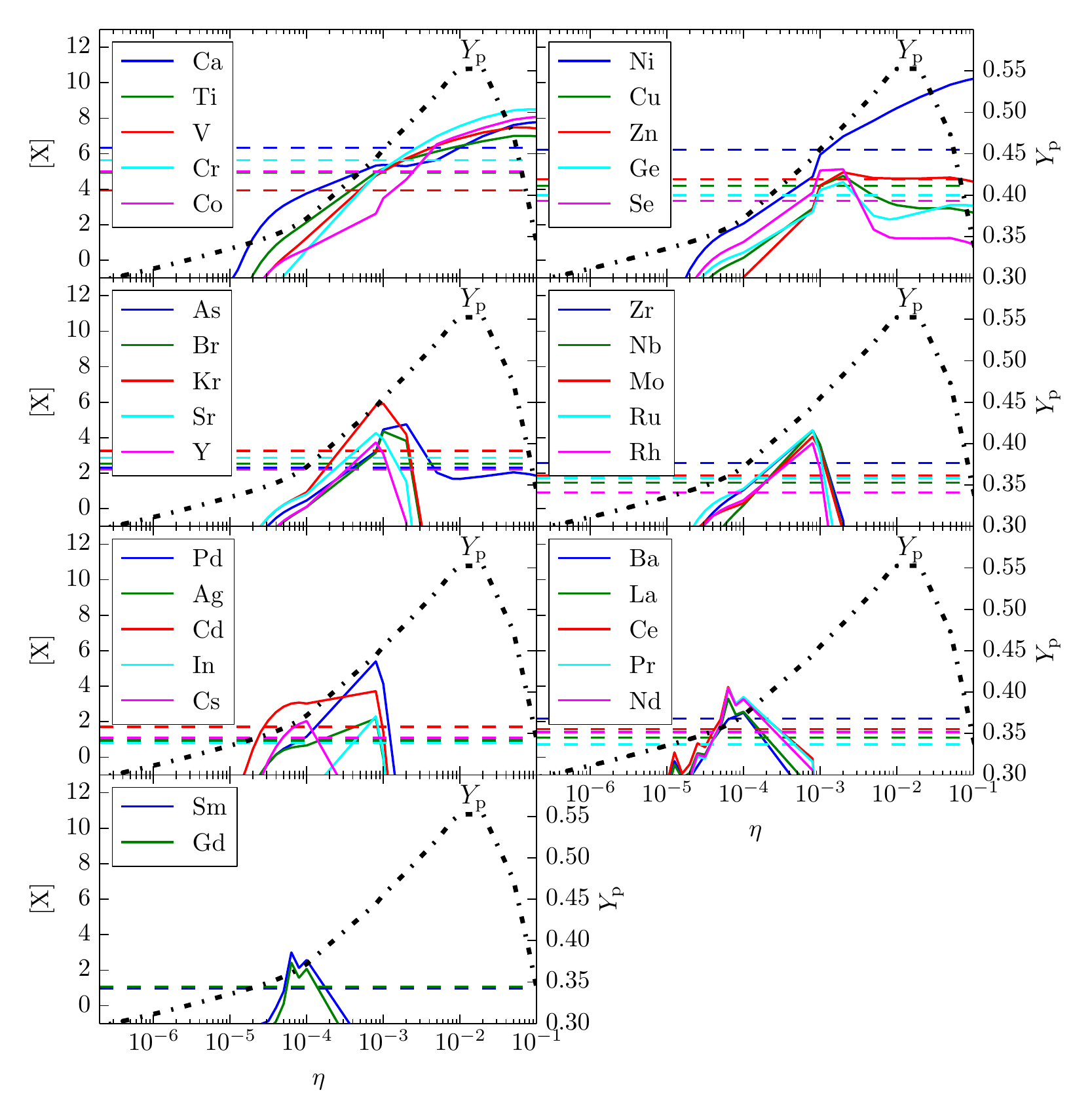}
		\vspace{-0.6cm}
		\caption{Abundance of nuclei computed by \texttt{AlterBBN} (plain curves) and compared to solar abundances of \cite{Asplund2009} (dashed curves, the error bars fit within their width). The $Y\mrm{p}$ abundance is shown as a dashed-dotted black line. In this figure we have retained only the elements (sorted by their atomic number $Z$) whose abundance exceeds the solar abundance somewhere in the $\eta$ range where $Y\mrm{p} \gtrsim 0.3$.\label{fig:abun_high_helium}}
	}
\end{figure*}

We focus here on the $^4$He abundance in globular star clusters. There is a long history of helium abundance determinations in these systems. The original motivation was that the oldest stars in the galaxies could provide clues for the primordial abundance. It was soon realized that the initial helium abundance affects several aspects of the evolution of globular cluster stars, including the  red giant  and  horizontal branch populations \cite{2004ARA&A..42..385G}. 
White dwarf properties are also affected \cite{2017A&A...602A..13C}.

Helium overabundances by as much as $\Delta Y\mrm{p}=0.1$ have been reported in massive globular clusters such as $\Omega$Cen \cite{2004ApJ...612L..25N} (see also \cite{Dupree2013} and references therein) and have more recently been shown to be correlated with abundance indicators of multiple stellar populations in globular clusters \cite{2018MNRAS.481.5098M}. Second generation recycling provides a means of enhancing the primordial helium abundance \cite{2018ARA&A..56...83B}. 

To summarise, there are  observational indications of enhanced helium in certain old stars. This is most likely due to mixing and evolutionary issues, and is expected to mask any possible primordial component. However should there be any surviving primordial abundance anomalies, they should be correlated in the same way that multiple population chromosome mapping has been performed in globular clusters \cite{2020MNRAS.493.6060S}.

Our hope is that the primordial inhomogeneous $\eta$ values indicative of a possible helium enhancement may reveal other abundance anomalies that we now discuss. With our program for evaluating high $\eta$ signatures we have indeed confirmed that there is a range of $\eta$ for which the $^4$He abundance is enhanced, namely $0.3 \lesssim Y_{\rm p} \lesssim 0.5\,,$ corresponding to $10^{-7} \lesssim \eta \lesssim 10^{-1}\,,$ as can be seen on Fig.~\ref{fig:abun_high_helium}. While a little overproduction of $^4$He occurs at $\eta \lesssim 10^{-7}$, it is not accompanied by the production of heavy elements.

This $\eta$ range of enhanced $^4$He corresponds to an overproduction (compared to solar values) of a variety of heavier elements, continuously dispersed in the range $A = 40 - 141$ ($Z = 20 - 59$). We plot in Fig.~\ref{fig:abun_high_helium} the predictions of the abundances of the elements (when summed upon their various isotopes) in the range of $\eta$ where the $^4$He abundance is $Y\mrm{p} \gtrsim 0.3$. We restricted the plots to elements for which the abundance is larger than the solar abundance for some values of $\eta$.

Very high values of $\eta \gtrsim 10^{-3}$ are associated with an overproduction of elements in the iron group $Z \sim 25$ (see also the upper panel of Fig.~\ref{fig:fixed_eta}) while somewhat lower values of the baryon-to-photon ratio $\eta \sim 10^{-4}$ are associated with an overproduction of neutron-capture elements in the $Z \sim 60$ region (see also the lower panel of Fig.~\ref{fig:fixed_eta}), some of which usually associated with $r$-process enhancement like Barium and Strontium (e.g. \cite{Ji2016}).

\begin{figure*}[t!]
    \centering
    \includegraphics[scale = 0.99]{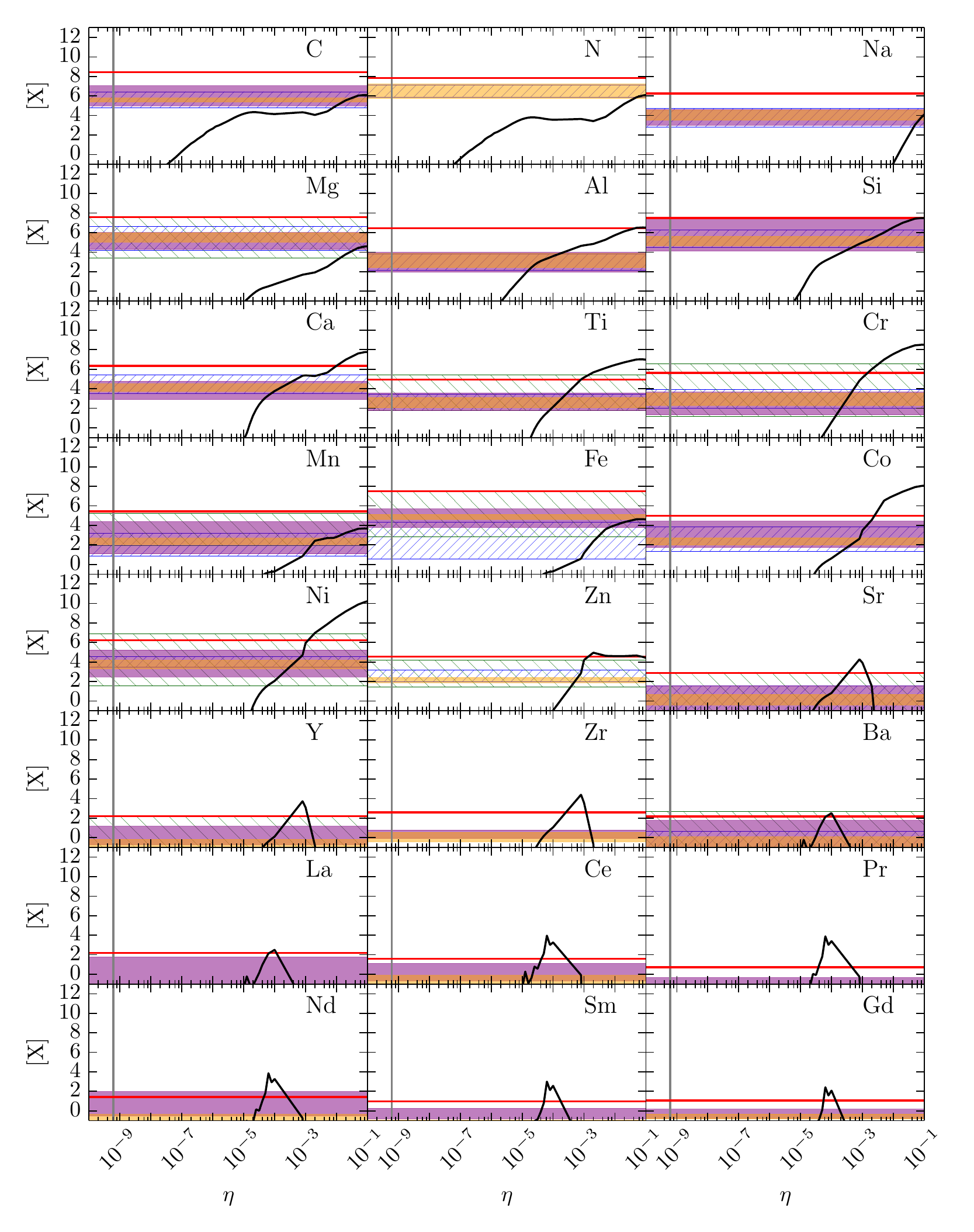}
    \caption{Comparison of \texttt{AlterBBN} data (black lines, summed over the isotopes) as functions of $\eta$ with metal-poor halo stars data \cite{Marino2019} (blue dashed areas), dwarf spheroidal faint galaxies data \cite{Reichert2020} (green dashed areas), and two $r$-process enhanced dwarf galaxies data: Reticulum II \cite{Ji2016} (purple areas) and Tucana III \cite{TucanaIII2019} (orange areas). We also represent the $1\sigma$ spread of the observational data. The solar abundances \cite{Asplund2009} are indicated as red lines. The standard BBN $1\sigma$ value for $\eta = (6.140\pm 0.058)\times 10^{-10}$ \cite{2020JCAP...03..010F} is represented as a grey vertical line. The error bars are smaller than the width of the lines.}
    \label{fig:comparisons}
\end{figure*}

\subsection{Comparison with metal-poor stars in the Galactic halo}
\label{sec:metal_poor}

The Galactic halo has proven to provide a remarkable environment for studying the oldest stars in the universe and for deciphering their evolution by chemical tagging \cite{Freeman:2002wq}. It is possible that such stars may contain signatures, hitherto overlooked, of high $\eta$ patches in the early universe that survived on the scales say of the smallest dwarf galaxies, the oldest galaxies in the universe.
Many of these systems later hierarchically merged into the halos of galaxies such as the Milky Way where they account for the enhanced $r$-process features observed  for $2-4\%$ of metal-poor halo stars \cite{2019ApJ...871..247B}.

The distinguishing characteristic of these metal-poor halo stars is that $\alpha/$Fe and other chemical signatures monitor the role of short time-scale core-collapse SNe enrichment \cite{2020IAUS..351...24I}. Hence they are a natural laboratory for our proposed signature of inhomogeneous BBN.

In this Section, we compare the abundances obtained with \texttt{AlterBBN} at high $\eta$ to some observed abundances in distant metal-poor stars \cite{Marino2019} (see Fig.~\ref{fig:comparisons}, blue dashed areas). We see that elements like Ni, Cr, Ca and Si, which are rare products in metal-poor (population II) stars, are produced in great quantities in the high-$\eta$ BBN scenario. This could give hints of observable signatures in  distant metal-poor stars. Normal stellar thermonuclear fusion and non-standard high-$\eta$ BBN do not give the same chemical products. In principle, these differences could also be used as a distinguishing characteristic of inhomogeneous BBN. 

\subsection{Comparison with metal-poor stars in dwarf spheroidal and ultrafaint galaxies}

Dwarf spheroidal galaxies are most likely the oldest stellar systems in the universe, and the ultrafaint dwarf spheroidals have metal-poor components that formed at least $10\,$Gyr ago and are enriched in $r$-process elements by some currently uncertain combination  
of  core-collapse supernovae  and neutron star mergers, with the latter contribution ranging from essentially  
all \cite{Komiya2016}
to some
\cite{Duggan2018} or only a minor fraction 
\cite{Cote2019}.
The core-collapse supernovae  contribute on a time-scale of $\sim 10^8\,$yr while  neutron star mergers 
on a much longer time-scale comparable to that of SNIa Fe enrichment of $\sim 4\times10^9\,$yr \cite{2020A&A...634L...2S}. Such a delayed bimodal enrichment history allows the possibility that some stars, the prompt-forming, extremely metal-poor component, avoided late enrichment, and any exotic primordial BBN signatures would be, at least relatively, chemically undiluted. 
%We however point out that the complexity of numerical simulations of SNe 

In this Section, we compare our predicted inhomogeneous BBN data with the abundances recently measured in 12 dwarf spheroidal or faint galaxies \cite{Reichert2020} (see Fig.~\ref{fig:comparisons}, green dashed areas). The Reticulum II galaxy was extracted from their data because it is treated separately, see below. We see that several elements could be detectable at $\eta \gtrsim 10^{-5}$, including Co, Ni, Sr, Y, Zr, Ce, Pr, Nd, Sm, Gd. Our model predicts relative abundance excesses for these at fixed values of $\eta$, which may well of course be diluted by mixing of $r$-process chemical enrichment.

\begin{figure}[t!]
    \centering
    \includegraphics[scale = 1]{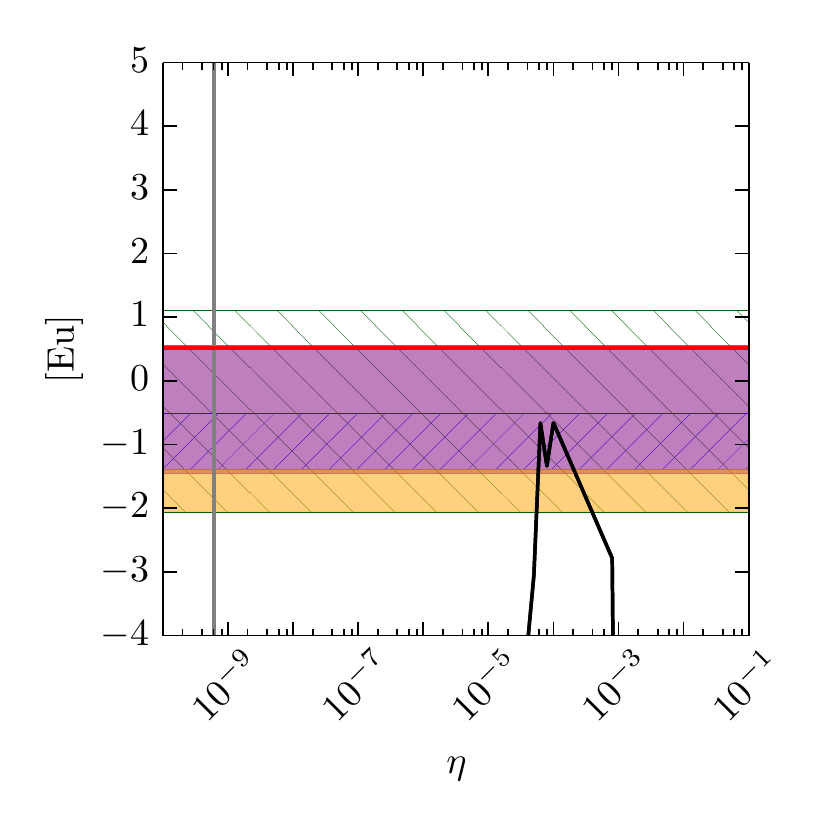}
    \caption{Comparison of \texttt{AlterBBN} Europium abundance (black line, summed over the isotopes) as a function of $\eta$ with metal-poor halo stars data \cite{Marino2019} (blue dashed areas), dwarf spheroidal faint galaxies data \cite{Reichert2020} (green dashed areas), and two $r$-process enhanced dwarf galaxies data: Reticulum II \cite{Ji2016} (purple areas) and Tucana III \cite{TucanaIII2019} (orange areas). We also represent the $1\sigma$ spread of the data. The solar abundance \cite{Asplund2009} is indicated as a red line. The standard BBN $1\sigma$ value for $\eta = (6.140\pm 0.058)\times 10^{-10}$ \cite{2020JCAP...03..010F} is represented as a grey vertical line. The error bars are smaller than the width of the lines.}
    \label{fig:europium}
\end{figure}

In particular, the nearby ultrafaint dwarf galaxies Reticulum II and Tucana III are fascinating laboratories for studying the evolution of and enrichment by the first stars in the universe. Unlike typical dwarfs, its \textit{r}-process history demonstrates a lack of complete mixing of supernovae ejecta with the ambient gas in its early gas-rich phase as compared to most other dwarfs \cite{2018ARNPS..68..237F}. In Fig.~\ref{fig:comparisons} we highlight the comparison of abundances obtained with \texttt{AlterBBN} at high $\eta$ to metal-poor stars observed in the dwarf galaxy Reticulum II \cite{Ji2016} (purple areas) and in Tucana III \cite{TucanaIII2019} (orange areas). In these galaxies, which are highly dark matter-dominated, there are hints of an enrichment in \textit{r}-process elements which in the absence of the most established astrophysical explanation, namely neutron star binaries \cite{2019MNRAS.490..296B}, could possibly correspond to a primordial component. A test of this would be to look for neutron-capture signatures that anti-correlate with the most common $r$-process product, Europium, and that is relatively unenhanced in our inhomogeneous BBN models\footnote{We had to specifically extract the Europium data from the complete set of \texttt{AlterBBN} data since its abundance does not exceed the lower cut-offs we had set.}, as highlighted in Fig.~\ref{fig:europium}. Europium suffers from an observable bias due to its very low abundance in metal-poor stars, unless it is enhanced as in Reticulum II or Tucana III, thus the comparison between different metal-poor stars populations may be flawed.

\section{Conclusions}
\label{sec:conclusion}

While the horizon size is of order $10^5\,\rm M_\odot$ at the onset of BBN, when the neutron abundance is frozen in at an epoch of $T \sim 1\,$MeV, Ref.~\cite{carr18} notes that extreme curvature fluctuations on horizon scales could form rare PBHs in this mass range provided the initial conditions are highly nongaussian, so that extreme perturbations that affect BBN only occur in rare inhomogeneous patches of the universe. Such scales are important as they provide PBHs of masses that are capable of seeding SMBHs observed at $z\lesssim 10$, when such seeding may be needed \cite{inayoshi}, and even could simultaneously accelerate galaxy formation. Mergers of these IMBHs will be detectable by LISA. We argue that a corollary, if confirmed as PBHs, for example by the redshift dependence of the merger rate, will be the possible but rare neutron-capture signatures in the oldest stars.

If sufficiently anomalous in terms of rare elements, the much later formation of stars in these regions may reflect primordial contamination by unusual abundance patterns. Of course this depends on the possible effects of radiation damping, not significant for isocurvature fluctuations, and of gas mixing throughout galactic evolution. Hence we suggest looking at either extremely metal-poor components of dwarf galaxies, which have mostly not undergone mergers, or the halo stars that most likely formed in now-merged dwarf building blocks. 

Any $r$-process enhancement is a signature of inefficient mixing at the epoch of the events, whether core-collapse SNe or neutron star mergers, that sourced the $r$-process elements. Hence we suggest targeting the metal-poor dwarf spheroidals Reticulum II and Tucana III as ideal candidates for our proposed signatures.

Another test of our hypothesis would be to look for neutron-capture signatures that anti-correlate with the most common $r$-process product, Europium, which is relatively unenhanced in our inhomogeneous BBN models, compared to other neutron-capture elements.

\section*{Acknowledgments}

We would like to thank Eva Grebel  and  Moritz Reichert for providing access to data on dwarf galaxy abundances,
and to Moritz Reichert for helpful comments on the manuscript. JS also acknowledges useful discussions with David Nataf and Rosemary Wyse on helium abundances.

\bibliographystyle{apsrev4-2}
\bibliography{biblio}% Produces the bibliography via BibTeX.

\end{document}